\documentclass{optica-article}

\journal{oe}

\usepackage{amsmath,booktabs}
\usepackage{tabularx}
\usepackage{array}
\usepackage{graphicx}
\usepackage{cancel}

\newcommand{\ud}{\mathrm{d}}
\newcommand{\D}{\partial}
\newcommand{\B}{\mathbf}
\newcommand{\tx}{\text}

\newcommand{\oo}[1]{\overline{\overline{#1}}}

\newcommand{\pati}[1]{\noindent{\textcolor{blue}{\textit{}}}\\}

\begin{document}

\title{Generalized FDTD Scheme for the Simulation of Electromagnetic Scattering in Moving Structures}

\author{Zo\'e-Lise Deck-L\'eger\authormark{1,*}, Amir Bahrami\authormark{2}, Zhiyu Li\authormark{{3,2}} and Christophe Caloz\authormark{2}}

\address{\authormark{1}Polytechnique Montréal, Canada, 
\authormark{2}KU Leuven, Belgium, \authormark{3}Xi'an Jiaotong University, China}

\email{\authormark{*}zoe-lise.deck-leger@polymtl.ca} 


\begin{abstract*}
Electromagnetic scattering in moving structures is a fundamental topic in physics and engineering. Yet, no general numerical solution to related problems has been reported to date. We introduce here a generalized FDTD scheme to remedy this deficiency. That scheme is an extension of the FDTD standard Yee cell and stencil that includes not only the usual, physical fields, but also auxiliary, unphysical fields allowing a straightforward application of moving boundary conditions. The proposed scheme is illustrated by four examples -- a moving interface, a moving slab, a moving crystal and a moving gradient -- with systematic validation against exact solutions.
\end{abstract*}



\section{Introduction}
Moving structures are pervasive in electromagnetics. They may involve both moving bodies, such stars or vehicles, and moving perturbations, such as fluid or elastic waves. The category of moving-body structures has a nearly 300-year history and has lead to the discovery of many fundamental physical phenomena, such as Bradley aberration~\cite{Bradley_1729}, Doppler frequency shifting~\cite{doppler1842}, Fizeau dragging~\cite{fizeau1851hypotheses}, relativity~\cite{einstein1905elektrodynamik}, magneto-electric coupling~\cite{Rontgen_1888}, medium bianisotropy~\cite{minkowski1908grundgleichungen} and gravity emulation~\cite{Leonhardt_PRA_1999}. The category of moving-perturbation structures is more recent and particularly amenable to practical applications; it produces effects such as parametric amplification~\cite{tien1958parametric,cassedy1963p1,cassedy1967p2}, nonreciprocal transmission~\cite{yu2009complete,chamanara2017isolation}, deflected reflection~\cite{deck2017superluminal,shalaev2019}, space-time frequency transitions~\cite{lampe1978interaction,chamanara2019simultaneous}, pseudo Fizeau dragging~\cite{huidobro2019,caloz2022GSTEM}, spatiotemporal bandgaps~\cite{biancalana2007dynamics,mattei2017field,sharabi_optica_2022} and accelerated
metamaterial bending~\cite{bahrami2022electrodynamics}.

Surprisingly, no general numerical tool is currently available to simulate electromagnetic moving structures, whether of the moving-body or moving-perturbation type. The development of such a tool would be highly desirable to simulate the plethora of moving structures mentioned above as well as future ones, particularly the noncanonical -- and typically more practical -- structures that do not admit analytical solutions. This issue is well-known. There have been only two workarounds in the litterature so far~\cite{Harfoush1990,Harfoush1989,iwamatsu2009,zheng2016,zhao_2018}, both based on the FDTD technique~\cite{amesnumerical,Taflove_2000}, probably selected for its natural incarnation of both spatial and temporal variations in Maxwell's equations. However, one of these approaches is restricted to non-penetrable objects~\cite{Harfoush1990,Harfoush1989}, while the other one implies cumbersome Lorentz frame transformations~\cite{iwamatsu2009,zheng2016,zhao_2018}.

We present here a novel, general and efficient FDTD scheme for simulating electromagnetic scattering in moving structures. That scheme, contrarily to that in~\cite{Harfoush1990,Harfoush1989}, also applies to penetrable media, allowing hence to handle gradient structures and metamaterials\cite{huidobro2019,deck2019uniform}, without requiring problematic numerical transformations, contrarily to that in~\cite{iwamatsu2009,zheng2016,zhao_2018}. This scheme is based on a generalized Yee cell and stencil, which include not only the usual, physical fields, but also auxiliary, unphysical fields that carry the velocity information, so as to automatically satisfy the moving boundary conditions.

\section{Types of Systems} 

\begin{figure}[h!]
    \centering
\includegraphics[width=0.6\columnwidth]{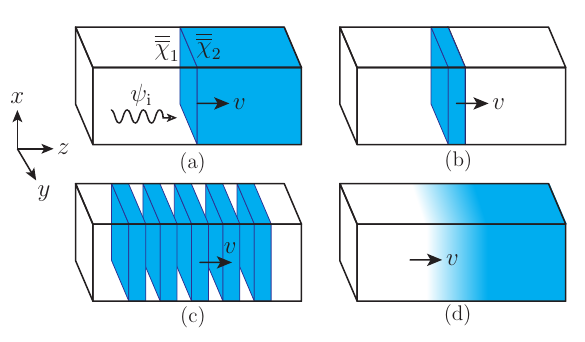}
    \caption{Types of moving structures, of either the moving-matter or moving-perturbation type, that can be simulated by the proposed generalized FDTD scheme. (a)~Simple interface. (b)~Slab. (c)~Bilayer crystal. (d)~Gradient.}
    \label{fig:geometry}
\end{figure}

The proposed method applies to both moving-matter and moving-perturbation media~\cite{caloz2022GSTEM}, the only difference between the two, in terms of modeling, being that in the former case, motion transforms media that are isotropic at rest into bianisotropic media~\cite{kong1968,bolotovski}, whereas in the latter case, motion alters the values of the permittivity or permeability parameters without affecting the isotropy of the medium~\cite{tsai1967wave,lampe1978interaction,caloz2019spacetime1}. Figure~\ref{fig:geometry} depicts some of these structures\footnote{In the case of moving-perturbation, these structures are essentially \emph{traveling-wave} modulation-type structures, while \emph{standing-wave} modulation-type structures, such as those studied in~\cite{milton2017field,sharabi_optica_2022}, which consist combinations of purely spatial and purely temporal interfaces can be handled by standard FDTD algorithm. Indeed, the FDTD algorithm naturally enforces continuity of the tangential $\B{E}$ and $\B{H}$ fields at stationary interfaces and continuity of $\B{D}$ and $\B{B}$ fields at temporal interfaces.}, which will be used for illustration and validation purposes in Sec.~\ref{sec:illutrations}: a simple interface~\cite{li2022total} [Fig.~\ref{fig:geometry}(a)], a slab~\cite{yee1966} [Fig.~\ref{fig:geometry}(b)], a bilayer crystal [Fig.~\ref{fig:geometry}(c)], which can be operated in the Bragg regime~\cite{cassedy1963p1} or in the metamaterial (homogeneous) regime~\cite{huidobro2019}, and a gradient~\cite{shvartsburg2005} [Fig~\ref{fig:geometry}(d)], all moving at a velocity $v$ in a direction selected as being $z$ in a Cartesian coordinate system. The different parts of these structures can naturally represent arbitrary bianisotropic media, with general tensor $\oo\chi=\left[\oo\epsilon,\oo\xi;\oo\zeta,\oo\mu\right]$, which may be all different from each other and may be arbitrarily truncated in the $xy$-plane. Note that all these structures can be seen as a succession of interfaces, including the gradient considering a decomposition in subwavelength slabs. For this reason, we shall focus on an interface in the next section, which deals with boundary conditions.

\section{Generalized Yee Cell}\label{sec:gen_Yee_cell}

The boundary conditions at a moving interface, whether associated to moving modulation or moving matter, read~\cite{pauli1981theory,bladel1984}
\begin{subequations}\label{eq:cont}
\begin{equation}\label{eq:Estarcont}
 \hat{\mathbf{n}}\times(\B{E}_2^*-\B{E}_1^*)=0,
\end{equation}
\begin{equation}\label{eq:Hstarcont}
  \hat{\mathbf{n}}\times(\B{H}_2^*-\B{H}_1^*)=\mathbf{J}_\tx{s},
\end{equation}
\begin{equation}\label{eq:Dcont}
     \hat{\mathbf{n}}\cdot(\B{D}_2-\B{D}_1)= \rho_\tx{s},
\end{equation}
\begin{equation}\label{eq:Bcont}
     \hat{\mathbf{n}}\cdot(\B{B}_2-\B{B}_1)= 0,
\end{equation}
\end{subequations}
with
\begin{equation}\label{eq:star}
    \B{E}^*=\B{E}+\mathbf{v}\times\B{B}
    \quad\text{and}\quad
    \B{H}^*=\B{H}-\mathbf{v}\times\B{D},
\end{equation}
where $1$ and $2$ label the media at the two sides the interface [as in Fig.~\ref{fig:geometry}(c)], $\B{E}$, $\B{H}$, $\B{D}$, $\B{B}$ are the usual electromagnetic fields, $\B{J}_\tx{s}$ and $\rho_\tx{s}$ are the usual surface current and charge densities, respectively, $\hat{\mathbf{n}}$ is the unit vector normal to the interface pointing towards medium~1, and $\B{v}$ is the velocity of the interface. Note that these conditions, as expected, reduce to the stationary boundary conditions for $\B{v}=0$, viz., $\hat{\mathbf{n}}\times(\B{E}_2-\B{E}_1)=0$, $\hat{\mathbf{n}}\times(\B{H}_2-\B{H}_1)=\B{
J}_\tx{s}$, $\hat{\mathbf{n}}\cdot(\B{D}_2-\B{D}_1)=\rho_\tx{s}$ and $\hat{\mathbf{n}}\cdot(\B{B}_2-\B{B}_1)=0$.

 The fields in Eq.~\eqref{eq:star} may decomposed into tangential and normal components with respect to the interface. The tangential components are
\begin{subequations}\label{eq:EHstar_comp}
\begin{equation}\label{eq:EHstart_tan}
\B{E}^*_\tx{tan}=\B{E}_\tx{tan}+\mathbf{v}\times\B{B}-(\hat{\mathbf{n}}\cdot(\mathbf{v}\times\B{B}))\hat{\mathbf{n}}\quad\text{and}\quad
\B{H}^*_\tx{tan}=\B{H}_\tx{tan}-\mathbf{v}\times\B{D}-(\hat{\mathbf{n}}\cdot(\mathbf{v}\times\B{D}))\hat{\mathbf{n}},
\end{equation}
while the normal components are
\begin{equation}\label{eq:EHstart_norm}
\B{E}^*_\tx{norm}=\B{E}_\tx{norm}+(\hat{\mathbf{n}}\cdot(\mathbf{v}\times\B{B}))\hat{\mathbf{n}}\quad\text{and}\quad
\B{H}^*_\tx{norm}=\B{H}_\tx{norm}+(\hat{\mathbf{n}}\cdot(\mathbf{v}\times\B{D}))\hat{\mathbf{n}}.
\end{equation}
\end{subequations}
In the case where the interface is perpendicular to the direction of motion, i.e., $\B{n}\|\B{v}$, the right-most terms in the right hand sides of Eqs.~\eqref{eq:EHstart_tan} and Eqs.~\eqref{eq:EHstart_norm} vanish. This is the scenario that is represented in Fig.~\ref{fig:geometry} and that will be assumed in the sequel of the paper\footnote{If we had $\B{n}\cancel{\|}\B{v}$, the interface would be slanted with respect to the motion direction. Related problems were treated analytically in~\cite{shiozawa1968general,censor1969scattering}}. The paper will further assume non-conducting interfaces, and hence $\B{J}_\tx{s}=\rho_\tx{s}=0$ in Eq.~\eqref{eq:cont}.

\begin{figure}[h!]
\centering
\includegraphics[width=0.5\textwidth]{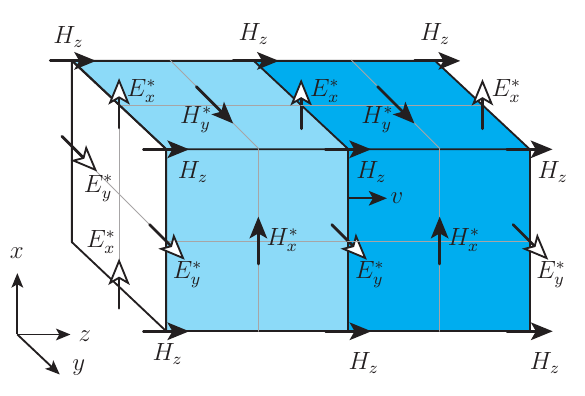}
\caption{Pair of generalized Yee cells, separated by a moving interface [Fig.~\ref{fig:geometry}(a)], with the generalization consisting in the substitution of the tangential physical fields $\B{E}$ and $\B{H}$ by the corresponding tangential auxiliary fields $\B{E}^*$ and $\B{H}^*$ in Eq.~\eqref{eq:star}.}
\label{fig:yee}
\end{figure}

 The dynamic boundary conditions~\eqref{eq:cont} differ from the static or stationary boundary equations only insofar as they involve the auxiliary fields $\B{E}_{1,2}^*$ and $\B{H}_{1,2}^*$ instead of the fields $\B{E}_{1,2}$ and $\B{H}_{1,2}$, whose tangential components include velocity-dependent extra terms [Eq.~\eqref{eq:EHstart_tan}]. This observation suggests the generalization of the standard Yee cell~\cite{yee1966} to a Yee cell with correspondingly modified $\B{E}_\tx{tan}$ and $\B{H}_\tx{tan}$ fields, without alteration of the conventional electric and magnetic field staggering arrangement. The result is shown in Fig.~\ref{fig:yee}, for a pair of generalized Yee cells separated by a moving interface. The so-defined generalized Yee cell straightforwardly allows to satisfy the field continuity conditions at a moving interface, i.e., Eqs.~\eqref{eq:Estarcont} and~\eqref{eq:Hstarcont}, by having the field components $E_x^*$, $E_y^*$ and $H_z$ uniquely defined at the permittivity interface of Fig.~\ref{fig:yee} (or $H_x^*$, $H_y^*$ and $E_z$ in the case of a corresponding magnetic interface).

 This simple generalization of the Yee cell is the core idea of the paper. This idea might a priori appear to be trivial, but it is not so trivial insofar as it implies a \emph{mixture of physical and unphysical fields}, namely, as seen in Eqs.~\eqref{eq:cont} and~\eqref{eq:EHstar_comp}, the physical fields $\B{D}$, $\B{B}$, $\B{E}_\mathrm{norm}$ and $\B{H}_\mathrm{norm}$ and the unphysical fields $\B{E}_\mathrm{tan}^*$ and $\B{H}_\mathrm{tan}^*$, which will also imply unphysical mixed-field electromagnetic equations. This strange situation is somewhat reminiscent to the unphysical anisotropic media appearing in Bérenger's Perfect Matched Layers (PMLs)~\cite{berenger1993} and, as the PMLs, it does not pose any fundamental problem. The corresponding unphysical Maxwell's equations constitutive relations, which will be derived and discretized in Sec.~\ref{sec:gen_scheme} for establishing the generalized FDTD update equations, are mathematical mappings of the physical Maxwell's equations via the simple changes of variable in Eqs.~\eqref{eq:EHstar_comp}, and the physical fields $\B{E}$ and $\B{H}$ can be found anytime from the auxiliary fields by simply inverting these relations.

\section{Generalized Scheme}\label{sec:gen_scheme}

 In this section, we shall derive the generalized FDTD scheme corresponding to the generalized Yee cell established in Sec.~\ref{sec:gen_Yee_cell}. This scheme must, of course, model the physics of the problem, and hence follow the usual Maxwell-Faraday and Maxwell-Ampère equations,
\begin{equation}\label{eq:Maxwell}
\frac{\D \B{B}}{\D t}=-\nabla\times\B{E} 
\quad\tx{and}\quad
\frac{\D \B{D}}{\D t}=\nabla\times\B{H},
\end{equation}
as well as the general constitutive relations,
\begin{equation}\label{eq:const_3d_regular}
        \B{D}=\oo\epsilon(\B{r}-\B{v}t)\cdot\B{E}+\oo{\xi}(\B{r}-\B{v}t)\cdot\B{H}
        \quad\tx{and}\quad
        \B{B}=\oo\zeta(\B{r}-\B{v}t)\cdot\B{E}+\oo\mu(\B{r}-\B{v}t)\cdot\B{H},
\end{equation}
whose bianisotropic form~\cite{kong2008theory} pertains to moving matter\footnote{In this case, for motion in the $z$ direction ($\B{v}=v\hat{\mathbf{z}}$), the tensors in Eq.~\eqref{eq:const_3d_regular} are~\cite{kong2008theory}
\begin{subequations}\label{eq:eqbianistens}
\begin{equation*}\label{eq:eqbianistensa}
\oo\epsilon=
\epsilon'\begin{bmatrix}
\alpha & 0 & 0\\
0 & \alpha & 0\\
0 & 0 & 1
\end{bmatrix}, \quad
\oo\mu=
\mu'\begin{bmatrix}
\alpha & 0 & 0\\
0 & \alpha & 0\\
0 & 0 & 1
\end{bmatrix}, \quad
\oo\xi=
\begin{bmatrix}
0 & \chi/c & 0\\
-\chi/c & 0 & 0\\
0& 0 & 0
\end{bmatrix}, \quad \oo\zeta=\oo\xi^T,
\end{equation*}
with
\begin{equation*}\label{eq:alpha_chi}
\alpha=\frac{1-\beta^2}{1-\beta^2{n'}^2}, \qquad \chi=\beta\frac{1-{n'}^2}{1-\beta^2{n'}^2}, \qquad \beta=v/c\qquad \tx{and} \qquad n'/c=\sqrt{\epsilon' \mu'},
\end{equation*}
\end{subequations}
where $\epsilon'$ and $\mu'$ are constant scalars that represent the permittivity and permeability of the medium at rest ($v=0$), which is assumed to be isotropic, nondispersive and linear.}
and which reduce to simple homoisotropic relations $\left(\oo{\epsilon}=\epsilon,~\oo{\mu}=\mu,~\oo{\xi}=\oo{\zeta}=0\right)$ in the case of moving modulation~\cite{tsai1967wave,lampe1978interaction,caloz2019spacetime1}.

 In order to apply the generalized Yee cell established in Sec.~\ref{sec:gen_Yee_cell}, we must express the fields $\B{E}$ and $\B{H}$ in Eqs.~\eqref{eq:Maxwell} and~\eqref{eq:const_3d_regular} in terms of the fields $\B{E}^*$ and $\B{H}^*$ using Eq.~\eqref{eq:star}. This results into the mixed-field Maxwell's equations
\begin{subequations}\label{eq:Maxwell_star}
\begin{equation}\label{eq:Faraday_star}
\frac{\D \B{B}}{\D t}=-\nabla\times\B{E}^*+\nabla\times(\B{v}\times\B{B}),
\end{equation}
\begin{equation}\label{eq:Ampere_star}
\frac{\D \B{D}}{\D t}=\nabla\times\B{H}^*+\nabla\times(\B{v}\times\B{D}),
\end{equation}
\end{subequations}
and constitutive relations
\begin{subequations}\label{eq:const_3d}
\begin{equation}
        \B{D}=\oo\epsilon(\B{r}-\B{v}t)\cdot\left(\B{E}^*-\B{v}\times\B{B}\right)+\oo{\xi}(\B{r}-\B{v}t)\cdot\left(\B{H}^*+\B{v}\times\B{D}\right),  
\end{equation}
\begin{equation}
        \B{B}=\oo\zeta(\B{r}-\B{v}t)\cdot\left(\B{E}^*-\B{v}\times\B{B}\right)+\oo\mu(\B{r}-\B{v}t)\cdot\left(\B{H}^*+\B{v}\times\B{D}\right).
\end{equation}
\end{subequations}

 From this point, we shall restrict our developments, for the sake of simplicity and without any loss of generality, to structures that are i)~based on moving modulation (as opposed to moving matter), ii)~one-dimensional (1D) and iii)~illuminated by a plane wave propagating in the direction of the modulation (i.e., normal to the interface(s)). Using a Cartesian coordinate system with $z$ corresponding to the direction of the modulation, the non-zero field components reduce to $E_x$ and $H_y$ and Eqs.~\eqref{eq:Maxwell_star} and~\eqref{eq:const_3d} become then
\begin{subequations}\label{eq:Maxwell_star_1d}
\begin{equation}\label{eq:Faraday_star_1d}
\frac{\D B_y}{\D t}=-\frac{\D E_x^*}{\D z}-v\frac{\D B_y}{\D z},
\end{equation}
\begin{equation}\label{eq:Ampere_star_1d}
\frac{\D D_x}{\D t}=-\frac{\D H_y^*}{\D z}-v\frac{\D D_x}{\D z},
\end{equation}
\end{subequations}
and
\begin{subequations}\label{eq:const_1d}
\begin{equation}\label{eq:D_const_1d}
        D_x=\epsilon(z-vt)\left(E^*_x+vB_y\right),
\end{equation}
\begin{equation}\label{eq:B_const_1d}
       B_y=\mu(z-vt)\left(H^*_y+vD_x\right),
\end{equation}
\end{subequations}
where we have used the fact that the mapping~\eqref{eq:star} preserves the polarization in the 1D case\footnote{To show this, we insert $\B{B}=-\int\left(\nabla\times\B{E}\right)\,\ud t$, from~\eqref{eq:Maxwell}, into~\eqref{eq:star}, which yields, using the vectorial identity $\B{A}\times(\nabla\times\B{B})=\nabla(\B{A}\cdot\B{B})-(\B{A}\cdot\nabla)\B{B}-(\B{B}\cdot\nabla)\B{A}-\B{B}\times(\nabla\times\B{A})$,
\begin{equation*}
    \B{E}^*=\B{E}-\smallint\left(\B{v}\times(\nabla\times \B{E})\right)\,\ud t=
    \B{E}-\smallint\left(\nabla(\mathbf{v}\cdot\B{E})-(\mathbf{v}\cdot\nabla)\B{E}-(\B{E}\cdot\nabla)\B{v}-\B{E}\times(\nabla\times \B{v}) \right)\,\ud t.
 \end{equation*}
The last two last terms in the last expression both vanish in the assumed regime of constant velocity. Given $\B{E}=E_x\hat{\B{x}}$ and $\B{v}=v\hat{\B{z}}$, the term with $\mathbf{v}\cdot\mathbf{E}$ also vanishes since $\mathbf{v}\perp\B{E}$. Finally, the term $(\mathbf{v}\cdot\nabla)\B{E}$ reduces to $\hat{\B{x}}\partial E_x/\partial z$ and is hence parallel to $\B{E}$. Thus, $\B{E}^*\|\B{E}$. It may be similarly shown that $\B{H}^*\parallel\B{H}$.}, so that only the star field components $E_x^*$ and $H_y^*$, corresponding to $E_x$ and $H_y$, are involved,

 We can now proceed to the discretization of Eqs.~\eqref{eq:Maxwell_star_1d} and~\eqref{eq:const_1d}. The static ($v=0$) parts can be discretized in the usual fashion~\cite{Taflove_2000}, which yields
\begin{subequations}\label{eq:discretize}
\begin{equation}\label{eq:Farad_discr}
B_y|^{n}_{k+\frac{1}{2}}=B_y|^{n-1}_{k+\frac{1}{2}}
-S\left( E^*_x|^{n-\frac{1}{2}}_{k+1}- E^*_x|^{n-\frac{1}{2}}_k\right)
-v\Delta t\frac{\D B_y}{\D z},
\end{equation}
\begin{equation}\label{eq:Ampere_discr}
 D_x|^{n+\frac{1}{2}}_k=D_x|^{n-\frac{1}{2}}_k
-S\left(H^*_y|^{n}_{k+\frac{1}{2}}-H^*_y|^{n}_{k-\frac{1}{2}}\right)-v\Delta t\frac{\D D_x}{\D z},
\end{equation}
and
\begin{equation}\label{eq:Dconst_discr}
E^*_x|_k^{n+\frac{1}{2}}=\frac{D_x|_k^{n+\frac{1}{2}}}{\epsilon|_k^{n+\frac{1}{2}}}-vB_y,
\end{equation}
\begin{equation}\label{eq:Bconst_discr}   
H^*_y|_{k+\frac{1}{2}}^{n}=\frac{B_y|_{k+\frac{1}{2}}^{n}}{\mu|_{k+\frac{1}{2}}^{n}}-vD_x,
\end{equation}
\end{subequations}
where $S=\Delta t/\Delta z$.

 In contrast, the dynamic ($v\neq 0$) terms, which have not been discretized yet in Eqs.~\eqref{eq:discretize}, require a special treatment for the numerical stability and minimal dispersion of the overall scheme. We empirically found that the discretization choices given in Tab.~\ref{tab:discretization} are appropriate in these regards. Note that these schemes are different for positive velocities, i.e., $\B{v}\|\B{\hat{z}}$ or $v>0$ and negative velocities, $\B{v}\|-\B{\hat{z}}$ or $v<0$.
\begin{table}
\abovedisplayskip=0pt
\belowdisplayskip=0pt
\noindent\begin{tabular*}{\textwidth}{@{\extracolsep{\fill}}c cc c c}
  &  Positive velocity ($v>0$) && Negative velocity ($v<0$) &\\
\end{tabular*}
\rule{\textwidth}{\heavyrulewidth}
\begin{subequations}\label{eq:dB}   
\noindent\begin{tabularx}{\textwidth}{XX}
  \begin{equation}\label{eq:dBa}
  \dfrac{\D B_y}{\D z}=\frac{B_{y}|^{n-1}_{k+\frac{1}{2}}-B_{y}|^{n-1}_{k-\frac{1}{2}}}{\Delta z}
  \end{equation} &
  \begin{equation}\label{eq:dBb}
 \dfrac{\D B_y}{\D z}=\frac{B_{y}|^{n-1}_{k+\frac{3}{2}}-B_{y}|^{n-1}_{k+\frac{1}{2}}}{\Delta z}
  \end{equation} 
\end{tabularx}
\end{subequations}
\begin{subequations}\label{eq:dD}   
\noindent\begin{tabularx}{\textwidth}{XX}
  \begin{equation}\label{eq:dDa} 
  \dfrac{\D D_x}{\D z}=\frac{D_{x}|^{n-1/2}_{k}-D_{x}|^{n-1/2}_{k-1}}{\Delta z}
  \end{equation} &
  \begin{equation}\label{eq:dDb} 
 \dfrac{\D D_x}{\D z}=\frac{D_{x}|^{n-1/2}_{k+1}-D_{x}|^{n-1/2}_{k}}{\Delta z}
  \end{equation} 
\end{tabularx}
\end{subequations}
\begin{subequations}\label{eq:B} 
\noindent\begin{tabularx}{\textwidth}{XX}
  \begin{equation}\label{eq:Ba}  
 B_y=\frac{B_{y}|^{n}_{k-1/2}+B_{y}|^{n}_{k-3/2}}{2}
  \end{equation} &
  \begin{equation}\label{eq:Bb}
  B_y=\frac{B_{y}|^{n}_{k+3/2}+B_{y}|^{n}_{k+1/2}}{2}
  \end{equation} 
\end{tabularx}
\end{subequations}
\begin{subequations}\label{eq:D}   
\noindent\begin{tabularx}{\textwidth}{XX}
  \begin{equation}\label{eq:Da} 
D_x=\frac{D_x|_{k+1}^{n-1/2}+D_x|_{k}^{n-1/2}}{2}
  \end{equation} &  
  \begin{equation}\label{eq:Db}
D_x=\frac{D_x|_{k+1}^{n-1/2}+D_x|_{k}^{n-1/2}}{2}
  \end{equation} 
\end{tabularx}
\end{subequations}
\rule{\textwidth}{\heavyrulewidth}
\caption{Discretization of the dynamic terms in Eqs.~\eqref{eq:discretize}.}\label{tab:discretization}
\end{table}

\begin{figure}[h!]
\centering
\includegraphics[width=1\textwidth]{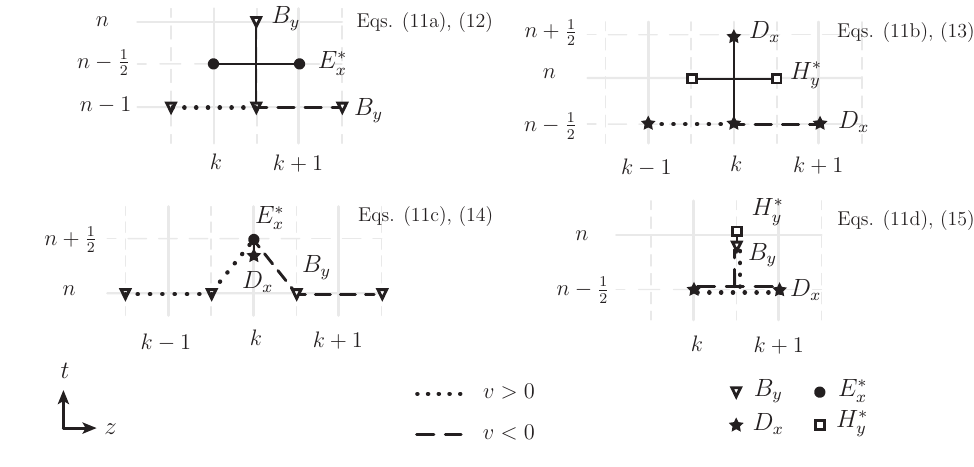}
\caption{FDTD stencil corresponding to the FDTD scheme in Eqs.~\eqref{eq:discretize} with Eqs.~\eqref{eq:dB}-\eqref{eq:D}.}
\label{fig:stencil}
\end{figure}

Figure~\ref{fig:stencil} shows the stencil of the overall proposed generalized FDTD scheme, which provides a handy pictorial representation of Eqs.~\eqref{eq:discretize} with Eqs.~\eqref{eq:dB}-\eqref{eq:D}.

\section{Stability}\label{sec:stability}

 In this section, we demonstrate the stability of the generalized scheme presented in Sec.~\ref{sec:gen_scheme} using von Neumann's approach~\cite{vonneumann1949}. This approach consists in inserting plane-wave test fields with the form
\setcounter{equation}{13}
\begin{equation}\label{eq:PW_test_field}
\Psi|_k^n=\Psi_0\xi^k\zeta^n,
\quad\tx{where}\quad
\xi=e^{ik_z \Delta z}
\quad\tx{and}\quad
\zeta=e^{-\alpha \Delta t}
\end{equation}
into the update equations and determining the condition under which the $\zeta(\xi)$ solution of the resulting matrix system is smaller than one -- but as close as possible to one to avoid numerical attenuation\footnote{In Eq.~\eqref{eq:PW_test_field}, one must be carefull not to confuse $k_z$, the $z$-component of the wavenumber, with $k$, the spatial index appearing in the FDTD update equations. No related ambiguity should exist in the paper insofar as the two quantities are always distinguishable from each other by the presence or absence of the $z$ subscript.}.

 Inserting the test fields of the above form into Eqs.~\eqref{eq:discretize} with Eqs.~\eqref{eq:dB}(a)-\eqref{eq:D}(a) or \eqref{eq:dB}(b)-\eqref{eq:D}(b), setting the determinant of the matrix associated with the resulting homogeneous system of equations to zero, and finding the $\zeta$ roots of the corresponding characteristic polynomial yields (see details in Appendix~\ref{sec:app_stability})
\begin{equation}\label{eq:zeta_stability}
\begin{split}
\zeta_\pm
&=\frac{\xi^{-3}}{4 n^2} \left[Sd\xi^4+4\left( b- 4S^2\right)\xi^3+S\left(3 \beta  n^2+2 S\right)\xi^2-\beta  n^2 S\xi \right]\\
&\quad\pm \frac{\xi^{-2}}{4 n^2}(1-\xi)S\sqrt{\xi^4 d^2+4  \left(b-2S^2\right)\xi^3+4 (n^4\beta^2-6n^2S\beta-2S^2)\xi^2 -4\beta ^2 n^2S\xi}
\end{split},
\end{equation}
where $\beta=v/c$, $\xi=e^{ik_z \Delta z}$, $b=n^2(4-3\beta S)$ and $d=2S+\beta n^2$. Setting $\zeta=1$ in Eq.~\eqref{eq:zeta_stability} provides the stability threshold of the scheme, with stability achieved for $\zeta<1$ (and unstability for $\zeta>1$).

 The fact that the stability analysis admits a \emph{double solution}, $\zeta_\pm$ [Eq.~\eqref{eq:zeta_stability}], may a priori appear suprising, given that $\zeta$ solutions are usually unique. However, this can be understood by realizing that waves that are \emph{co-moving and contra-moving} with respect to the modulation see different numerical ``landscapes''. Specifically, if $v>0$, then the scheme uses Eq.~\eqref{eq:dBa}-\eqref{eq:Da} [first column in Tab.~\ref{tab:discretization}] for the dynamic parts, which correspond to the dotted segments (with the dashed segment removed) in Fig.~\ref{fig:stencil}; the resulting stencils for the first three equations [Eqs.~\eqref{eq:Farad_discr}-\eqref{eq:Dconst_discr} with Eqs.~\eqref{eq:dBa}-\eqref{eq:Ba}] are clearly asymmetric with respect the $z$ direction and therefore, although the stencil for the fourth equation [Eq.~\eqref{eq:Bconst_discr} with Eq.~\eqref{eq:Da}] is symmetric, the stencil is \emph{overall asymmetric} with respect to $z$, so that co-moving and contra-moving waves will see different stencils. A similar conclusion holds for the case $v<0$, so we expect that, for a given choice of the modulation direction, $\zeta_+$ will correspond to one of the two (co-moving or contra-moving) regime, while $\zeta_-$ will correspond to the other regime.

 The two regimes may be identified by comparing the signs of the spatial and temporal phases of the test plane-wave in Eq.~\eqref{eq:PW_test_field}, with the spatial phase and temporal phase residing in $\xi$ and $\zeta_\pm$, respectively. For this purpose, let us consider a specific numerical position and time, say for simplicity $k=1$ and $n=1$, and analyze the test plane-wave for the two possible propagation directions. We have, upon writing $\alpha_\pm=\omega_{\tx{i}\pm}\pm i\omega_{\tx{r}\pm}$ in $\zeta_\pm$ (with $\omega_{\tx{i}\pm}>0$ assuming stability),
\begin{equation}\label{eq:PW_anal}
\begin{split}
    \frac{\Psi|_{k=1}^{n=1}}{\Psi_0}=\xi\zeta_\pm
    &=e^{ik_z\Delta z-\alpha_\pm\Delta t}
    =e^{ik_z\Delta z-(\omega_{\tx{i}\pm}\pm i\omega_{\tx{r}\pm})\Delta t}\\
    &=e^{-\omega_{\tx{i}\pm}\Delta t}e^{i(k_z\Delta z\mp\omega_{\tx{r}\pm}\Delta t)}
    =e^{-\omega_{\tx{i}\pm}\Delta t}e^{ik_z\Delta z}e^{\mp i\omega_{\tx{r}\pm}\Delta t}\\
    &=e^{ik_z \Delta z}e^{-\omega_{\tx{i}\pm}\Delta t}\left[\cos(\omega_{\tx{r}\pm}\Delta t)\mp i\sin(\omega_{\tx{r}\pm} \Delta t)\right]=\xi\left(\zeta_{\tx{r}\pm}\mp i\zeta_{\tx{i}\pm}\right).
    \end{split}
\end{equation}
In  these relations, the first expression of the second line reveals that the upper sign, which is associated with $\zeta_+$, corresponds to a forward wave, propagating in the $+z$ direction, while the lower sign, which is associated with $\zeta_-$, corresponds to a backward wave, so that $\zeta_+$ and $\zeta_-$ respectively correspond to the co-moving and contra-moving regimes for $v>0$ (and conversely for $v<0$). Note that the scheme must be stable for both waves, since typical problems involve both of them\footnote{This is even true for a simple mismatched moving interface, where the incident and transmitted waves are co-moving and the reflected wave is contra-moving if the incident wave propagates in the direction of the modulation.}. The stability criterion should then be taken as that corresponding to the more restrictive of the two $\zeta$ solutions.

We shall now show that the $\pm$ signs in Eq.~\eqref{eq:PW_anal} correspond to $\pm$ signs in Eq.~\eqref{eq:zeta_stability}. We have to do this numerically because we have not been able to separate the real and imaginary parts of $\zeta$ in Eq.~\eqref{eq:zeta_stability} as we did in Eq.~\eqref{eq:PW_anal}. Inserting the parameters $S=0.5$, $\beta=0.3$, $\epsilon=4$, $\mu=1$ and $\Delta z k_z=2\pi /5$ into Eq.~\eqref{eq:zeta_stability} yields $\zeta_+=0.925-i 0.33$ and $\zeta_-=0.917+i 0.23$, which is indeed consistent with Eq.~\eqref{eq:PW_anal}, and hence confirms the consistency of the $\pm$ indices of $\zeta$ in Eqs.~\eqref{eq:zeta_stability} and~\eqref{eq:PW_anal}. Incidentally, the magnitudes of the $\zeta$'s, as apparent in the last equation of Eqs.~\eqref{eq:PW_test_field}, represent the levels of the wave, which are here $|\zeta_+|=0.98$ and $|\zeta_-|=0.95$, which reveals that the co-moving wave is less attenuated than the contra-moving wave.

\begin{figure}[h!]
    \centering
    \includegraphics[width=1\columnwidth]{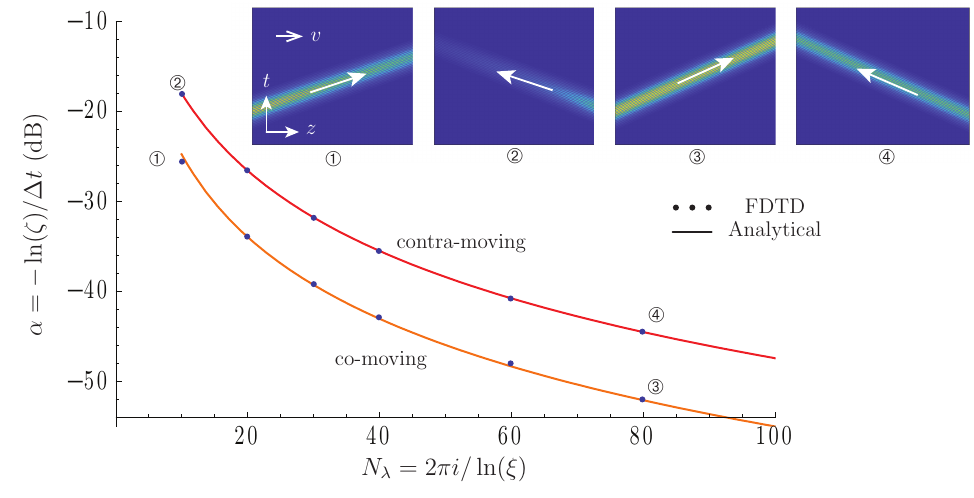}
    \caption{Attenuation rate versus meshing density obtained by Eq.~\eqref{eq:zeta_stability} and by FDTD simulation (taking field ratios at different times), for $S=0.5$, $v=0.3c$ ($v>0$, first column in Tab.~\ref{tab:discretization}), $\epsilon=1$ and $\mu=1$. The insets show FDTD-simulated space-time evolution of a $3\lambda$-wide Gaussian pulse modulated by a harmonic wave of wavelength $\lambda=2\pi/k_z$ and period $T=2\pi/\omega$, over a distance of $20\lambda$ and time of $45T$, for the mesh densities $N_\lambda=10$ and $N_\lambda=80$.}
    \label{fig:stability}
\end{figure}

 Figure~\ref{fig:stability} plots the attenuation rate $\alpha=\ln(\zeta)/\Delta t$ versus the meshing density $N_\lambda=\lambda_0/\Delta z=2\pi/(k_z\Delta z)=i 2\pi/\ln(\xi)$ obtained by Eq.~\eqref{eq:zeta_stability} and by FDTD simulations, along with FDTD-simulated space-time field evolutions for two meshing densities, for the case $v>0$ [corresponding to the stencil with Eqs.~\eqref{eq:dBa}-\eqref{eq:Da}]. The FDTD attenuation results closesly match the attenuation results predicted by the analytical stability analysis, with distinct levels corresponding, as expected, to the co-moving and contra-moving regimes. Here, with the choice $v>0$, the appropriate (more restrictive) criterion corresponds to that for the co-moving waves ($\zeta_+$), whereby the contra-moving ($\zeta_-$) waves will experience an extra level of numerical attenuation compared to what would be required if they were alone. Symmetrically identical results (not shown) are obtained for the case $v>0$, corresponding to the stencil with Eqs.~\eqref{eq:dBb}-\eqref{eq:Db}. 

\section{Illustrative and Validating Examples}\label{sec:illutrations}

\begin{figure}[h!]
    \centering
    \includegraphics[width=1\columnwidth]{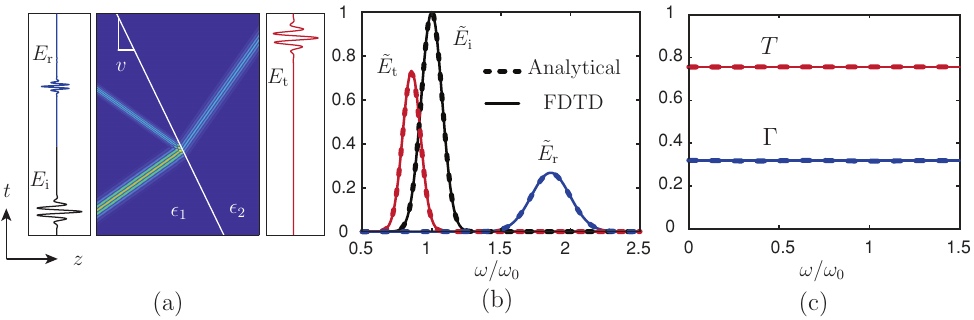}
    \caption{Results for a contra-moving interface [Fig.~\ref{fig:geometry}(a)], for the physical parameters $v=-0.3c$, $\epsilon_1=2$, $\epsilon_2=4$ and $\mu=1$, and numerical parameters $S=0.2$ and $\Delta z=\lambda_0/150$. The analytical solutions are given in Appendix~\ref{sec:analytical_interface}. (a)~Space-time evolution of the electric field for a modulated Gaussian pulse ($E_\tx{i}=e^{-i\omega_\tx{i} t}e^{-(t/\tau_\tx{i})^2}$), obtained by combining spatial waveforms at successive times, with input and output temporal waveforms plotted on the sides. (b)~Spectrum of the input and output fields, obtained by Fourier transforming the waveforms sampled at the input and output positions in~(a). (c)~Transmission and reflection coefficients, obtained by taking the ratios of the Fourier transforms of the appropriate fields for a short (wide-band) incident pulse.}
    \label{fig:results_interface}
\end{figure}

\begin{figure}[h!]
    \centering
    \includegraphics[width=1\columnwidth]{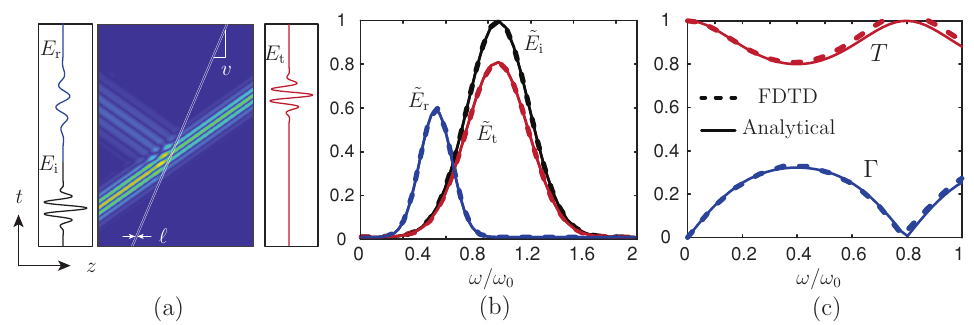}
    \caption{Same as in Fig.~\ref{fig:results_interface} but for a co-moving slab [Fig.~\ref{fig:geometry}(b)], with velocity $v=0.3c$, $\epsilon_2=4$ and length $\ell=\frac{\lambda_0}{4n_2}\frac{1+n_2\beta}{1-n_1\beta}$, corresponding to a space-time quarter-wave slab~\cite{deck2019uniform}.}
    \label{fig:results_slab}
\end{figure}

\begin{figure}[h!]
    \centering
    \includegraphics[width=1\columnwidth]{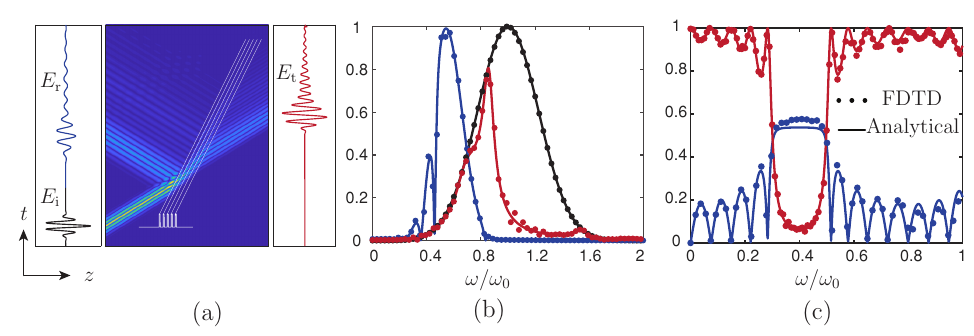}
    \caption{Same as in Fig.~\ref{fig:results_interface} but for a co-moving bilayer crystal [Fig.~\ref{fig:geometry}(c)], consiting of 5 unit-cells each made of two space-time quarter-wave slabs $\epsilon_1=1$, $\epsilon_2=4$, $\ell_1=\frac{\lambda_0}{4n_1}\frac{1+n_1\beta}{1-n_2\beta}$, $\ell_2=\frac{\lambda_0}{4n_2}\frac{1+n_2\beta}{1-n_1\beta}$,  }
    \label{fig:results_crystal_co}
\end{figure}

\begin{figure}[h!]
    \centering
    \includegraphics[width=1\columnwidth]{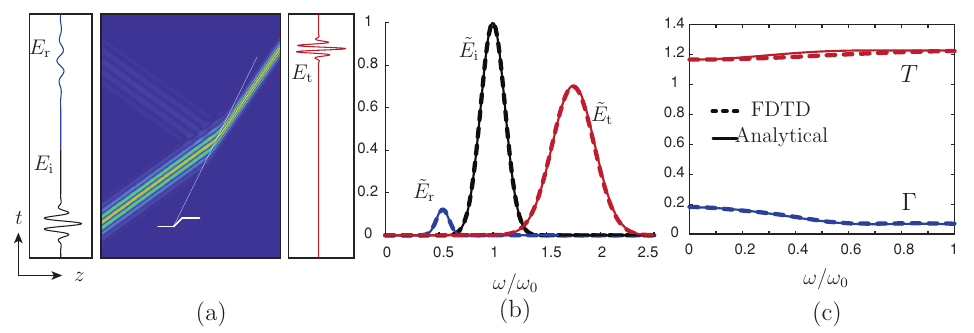}
    \caption{Same as in Fig.~\ref{fig:results_interface} but for a co-moving gradient [Fig.~\ref{fig:geometry}(d)], corresponding to a linear increase of the permittivity from $\epsilon_1$ to $\epsilon_2$ over a length of $\ell=\lambda_0/2$.}
    \label{fig:results_gradient}
\end{figure}

 Figures~\ref{fig:results_interface},~\ref{fig:results_slab},~\ref{fig:results_crystal_co} and~\ref{fig:results_gradient} provide illustrative and validating examples, corresponding to Figs.~\ref{fig:geometry}(a),~(b),~(c) and~(d), for the problems of electromagnetic scattering at a moving interface, a moving slab, a moving cystal and a moving gradient, respectively.

 In all the cases (Figs.~\ref{fig:results_interface},~\ref{fig:results_slab},~\ref{fig:results_crystal_co} and~\ref{fig:results_gradient}) and for all the quantities (incident, reflected and transmitted electric fields, and reflection and transmission coefficients), excellent agreement is observed between the numerical results of the proposed generalized FDTD scheme and the analytical results (given in Appendix~\ref{sec:analytical}), with minor observable discrepancies in the scattering coefficients, which are explained by the limited bandwidth of the test pulse.

 The physics of the dynamic scattering observed in Figs.~\ref{fig:results_interface},~\ref{fig:results_slab},~\ref{fig:results_crystal_co} and~\ref{fig:results_gradient} involves reflection Doppler frequency shifting, transmission index contrast frequency shifting, co-moving attenuation, contra-moving amplification, multiple space-time scattering and space-time stopbands. All these effects have been described elsewhere~\cite{deck2019uniform,caloz2019spacetime1,caloz2019spacetime2} and are therefore not discussed here.

\section{Conclusion}
 We have presented a generalized FDTD scheme that can simulate electromagnetic scattering in a great diversity of moving-medium problems. This scheme, using a combination of physical and auxiliary fields to satisfy moving boundary conditions in a natural fashion, is both simple and powerful. It fills an fundamental and important gap in the toolbox of electromagnetic computational techniques and is hence expected to find wide applications, particularly in the emergent area of space-time metamaterials.

\section*{Appendices}

\appendix

\section{Stability Calculation}\label{sec:app_stability}

 This section derives the stability condition for the generalized FDTD scheme, presented in Sec.~\ref{sec:gen_scheme}, using von Neumann's approach~\cite{vonneumann1949}. We will therefore replace everywhere the electromagnetic fields by their plane-wave test counterpart, $\Psi|_k^n=\Psi_0\zeta^n e^{ik_z k\Delta z}$, with $\zeta=e^{-\alpha \Delta t}$. In addition, we will use the convenient parameter $S=\Delta t/\Delta z$ wherever appropriate.

We shall restrict our derivations to the positive-velocity regime ($v>0$), whose dynamic part corresponds to the first column of Tab.~\ref{tab:discretization}. The negative-velocity derivations ($v<0$), whose dynamic part corresponds to the second column of Tab.~\ref{tab:discretization}, can easily be obtained upon following similar steps.

Using the plane-wave test field substitution in Eqs.~\eqref{eq:Farad_discr} and~\eqref{eq:dBa}, and dividing the resulting expression by $\zeta^n$ and $e^{ik_z k\Delta z}$ yields
%
\begin{multline}
B_{y0} e^{ik_z\Delta z/2}=B_{y0} e^{ik_z\Delta z/2}\zeta^{-1}
-SE^*_{x0}\zeta^{-1/2}e^{ik_z\Delta z/2}\left(e^{ik_z\Delta z/2}-e^{-ik_z\Delta z/2}\right)\\
-vSB_{y0}\zeta^{-1}\left(e^{ik_z\Delta z/2}-e^{-ik_z\Delta z/2}\right),
\end{multline}
%
which may be alternatively written, upon multiplying by $\zeta e^{-ik_z \Delta z/2}$, forming sines, grouping terms and transferring everything to the left hand-side of the equality, as
\begin{equation}\label{eq:1_reorg}
E^*_{x0}\zeta^{1/2}2iS\sin(k_z\Delta z/2)+B_0\left(\zeta-1+ve^{-ik_z\Delta z/2}2iS\sin(k_z\Delta z/2)\right)=0.
\end{equation}

Similarly, Eq.~\eqref{eq:Ampere_discr} with Eq.~\eqref{eq:dDa} becomes
\begin{equation}
D_{x0}\zeta^{1/2}=D_{x0}\zeta^{-1/2}
-SH^*_{y0}\left(e^{ik_z\Delta z/2}-e^{-ik_z\Delta z/2} \right)
-vSD_{x0}\zeta^{-1/2}e^{-ik_z\Delta z/2}\left(e^{ik_z\Delta z/2}-e^{-ik_z\Delta z/2}\right),
\end{equation}
or, multiplying by $\zeta^{1/2}$, and doing the same next operations as for the previous equation,
\begin{equation}\label{eq:2_reorg}
D_{x0}\left(\zeta-1+2iSve^{-ik_z\Delta z/2}\sin(k_z\Delta z/2)\right)+
H_{y0}^*\zeta^{1/2}2iS\sin(k_z\Delta z/2)=0.
\end{equation}

Next, Eq.~\eqref{eq:Dconst_discr} with~\eqref{eq:Ba} becomes, after factoring out $e^{-ik_z\Delta z}$ from the last two terms to form a cosine, 
\begin{equation}
E^*_{x0}\zeta^{1/2}=D_{x0}\zeta^{1/2}/\epsilon 
-\frac{v}{2}B_{y0}e^{-ik_z\Delta z}(e^{ik_z\Delta z/2}+e^{-ik_z\Delta z/2}),
\end{equation}
or, multiplying by $\epsilon\zeta^{-1/2}$, forming cosines, etc.,
\begin{equation}\label{eq:3_reorg}
D_{x0}-\epsilon E^*_{x0}-\epsilon vB_{y0}\zeta^{-1/2}e^{-ik_z\Delta z}\cos(k_z\Delta z/2)=0.
\end{equation}

Finally, Eq.~\eqref{eq:Bconst_discr} with~\eqref{eq:Da} becomes, after factoring out $e^{-ik_z\Delta z/2}$ from the last terms to form a cosine, 
\begin{equation}
H^*_{y0}e^{ik_z\Delta z/2}=B_{y0}e^{ik_z\Delta z/2}/\mu-\frac{v}{2}\zeta^{-1/2}e^{ik_z\Delta z/2}D_{x0}(e^{ik_z\Delta z/2}+e^{-ik_z\Delta z/2}),
\end{equation}
or, multiplying by $-\mu e^{-ik_z\Delta z/2}$, etc.,
\begin{equation}\label{eq:4_reorg}
B_{y0}-\mu H^*_{y0}-\mu v\zeta^{-1/2}D_{x0}\cos(k_z\Delta z/2)=0.
\end{equation}

Equations~\eqref{eq:1_reorg},~\eqref{eq:2_reorg},~\eqref{eq:3_reorg} and~\eqref{eq:4_reorg} form a system of equations that can be put in the matrix form
\begin{equation}
    \begin{bmatrix}
        A  & 0 & 0 &B+vC\\
       0  &  A & B+vC& 0\\
       -\epsilon   & 0 & 1 & -\epsilon v De^{-ik_z\Delta z}\\
        0 & -\mu  & -\mu v D & 1\\        
\end{bmatrix}
\begin{bmatrix}
    E^*_{x0}\\
    H^*_{y0}\\
    D_{x0}\\
    B_{y0}\\
\end{bmatrix}=0,
\end{equation}
where 
\begin{subequations}\label{eq:ABCD}  
\begin{equation}
 A=2i\zeta^{1/2}S\sin(k_z\Delta z/2),
\end{equation}
\begin{equation}
 B=\zeta-1,
\end{equation}
\begin{equation}
C=2iSe^{-ik_z\Delta z/2}\sin(k_z\Delta z/2),
\end{equation}
\begin{equation}
D=\zeta^{-1/2}\cos(k_z\Delta z/2).
\end{equation}
\end{subequations}
Setting the determinant of this homogeneous system to zero to find a nontrivial solution for $\zeta$ leads to the characteristic polynomial
\begin{equation}
\begin{split}\label{eq:zeta_polynom}
   \zeta^2 &+\frac{\zeta}{2n^2}\left(4S^2-n^2(4-3\beta S)-4S(S+n^2\beta)\cos(k_z\Delta z)\right)\\&+\frac{\zeta S\beta}{2}\left(\cos(2k_z\Delta z)+2i\sin(k_z\Delta z)-i\sin(2k_z\Delta z)\right)\\&+e^{-ik_z\Delta z}[1-S\beta(1-\cos(k_z\Delta z))][S\beta+(S\beta -1)\cos(k_z\Delta z)-i\sin(k_z\Delta z)]=0.
\end{split}
\end{equation}
This is a polynominal of the second order in $\zeta$ whose solution is given by Eq.~\eqref{eq:zeta_stability}. The corresponding polynomial and solutions turn out to be exactly identical for the negative-velocity regime, although the related derivation involves distinct equations.

\section{Analytical Solutions}\label{sec:analytical}
 This section provides analytical solutions that are used for validation of the illustrative examples given in Sec.~\ref{sec:illutrations}. The solutions in Secs.~\ref{sec:analytical_interface} and~\ref{sec:analytical_slab} are closed-form solutions for the canonical problems of an interface and a slab, corresponding to Figs.~\ref{fig:geometry}(a) and~\ref{fig:geometry}(b), respectively, while the solutions in Sec.~\ref{sec:analytical_tmm} are transfer-matrix method solutions for the more complex problems of a bilayer crystal and a gradient, corresponding to Figs.~\ref{fig:geometry}(c)~\ref{fig:geometry}(d), respectively.

\subsection{General Formulas}
 In all cases, the incident field is a modulated Gaussian pulse, with temporal and spectral profiles
\begin{equation}\label{eq:spectrum_incident}
     \psi_\tx{i}(t)=e^{-i\omega_\tx{i}t}e^{-(t/\tau_\tx{i})^2}
     \quad\tx{and}\quad
     \tilde\psi_\tx{i}(\omega)=\frac{\tau_\tx{i}}{4\sqrt{\pi}}e^{\tau_\tx{i}^2(\omega-\omega_\tx{i})/4},
\end{equation}
where $\tau_\tx{i}$ is the full-width at half-maximum duration divided by $2\sqrt{\ln 2}$ and $\omega_\tx{i}$ is the modulation frequency. The spectra of the reflected and transmitted pulses are obtained as
\begin{equation}\label{eq:spectrum_scat_interface}
 \tilde\psi_\tx{t,r}(\omega)=\{T(a_\tx{t}\omega),\Gamma(a_\tx{r}\omega)\} \frac{\tau_\tx{i}}{4\sqrt{\pi}a_\tx{t,r}}e^{\tau_\tx{i}^2(\omega-a_\tx{t,r}\omega_\tx{i})^2/(4a_\tx{t,r})},
\end{equation}
where $T$ and $\Gamma$ are the transmission and reflection coefficients, and $a_\tx{t,r}$ are the transmission and reflection compression or expansion factors. This factors are given by
\begin{equation}\label{eq:comp_exp}
    a_\tx{t}=\frac{1-n_\tx{in}v/c}{1-n_\tx{out}v/c}
    \quad\tx{and}\quad 
    a_\tx{r}=\frac{1-n_\tx{in}v/c}{1+n_\tx{in}v/c},
\end{equation}
where $n_\tx{in}$ is the refractive index of the medium in which the incident and reflected waves propagate, and $n_\tx{out}$ is the refractive index of the medium in which the transmitted wave propagates\footnote{The equations~\eqref{eq:spectrum_scat_interface} and~\eqref{eq:comp_exp} may be derived by combining results in~\cite{deck2019uniform,caloz2019spacetime2} for monochromatic waves; they do not seem to have been published explicitely anywhere so far.}.

\subsection{Interface}\label{sec:analytical_interface}
The scattering coefficients for a modulated interface are found by applying the continuity conditions~\eqref{eq:Estarcont} and~\eqref{eq:Hstarcont}. They read~\cite{tsai1967wave,deck2019uniform,caloz2019spacetime2}, assuming wave incidence from medium 1~towards medium~2,
\begin{equation}\label{eq:coef_interface}
    T_{21}=\frac{2\eta_2}{\eta_1+\eta_2}\frac{1-n_1v/c}{1-n_2v/c}
    \quad\tx{and}\quad
    \Gamma_{121}=\frac{\eta_2-\eta_1}{\eta_1+\eta_2}\frac{1-n_1v/c}{1+n_1v/c},
\end{equation}
while the formulas for the reverse propagation direction are obtained by a simnple interchange of the subscripts. Note that the scattering coefficients here independent of the frequency.

\subsection{Slab}\label{sec:analytical_slab}
The scattering coefficients for a modulated slab may be found, as for a stationnary slab, either by the multiple-reflection method or the boundary-value method~\cite{Sommerfeldoptics}. This results in the scattering coefficients~\cite{deck2019uniform}
\begin{equation}\label{eq:coef_slab}
    T=\frac{T_{21}T_{12}e^{-i\omega_2^+n_2\ell}}{1-\Gamma_{212}\Gamma^-_{212}e^{-i(\omega_2^++\omega_2^-)n_2\ell}}\quad  \text{and} \quad\Gamma=\Gamma_{121}\frac{1-e^{-i(\omega_2^++\omega_2^-)n_2\ell}}{1-\Gamma_{212}\Gamma^-_{212}e^{i(\omega_2^++\omega_2^-)n_2\ell}}.
\end{equation}
These relations imply different frequencies for waves propagating forward ($\omega_2^+$) and backward ($\omega_2^-$), namely
\begin{equation}
    \omega_2^+=\frac{1-n_1v/c}{1-n_2v/c}\omega_\text{i}\quad \text{and} \quad\omega_2^-=\frac{1-n_1v/c}{1+n_2v/c}\omega_\text{i},
\end{equation}
where $T_{21}$ and $\Gamma_{121}$ are provided in Eq.~\eqref{eq:coef_interface}, $T_{12}$ and $\Gamma_{121}$ are obtained by interchanging $1$ and $2$ in Eq.~\eqref{eq:coef_interface}, and $\Gamma_{212}^-$ is obtained by interchanging $1$ and $2$ and changing the sign of $v$ in $\Gamma_{121}$~\eqref{eq:coef_interface}.

\subsection{Transfer-Matrix Method}\label{sec:analytical_tmm}
 The transfer-matrix method~\cite{tsai1967wave,biancalana2007dynamics,deck2019uniform} allows to compute electromagnetic scattering for any traveling-wave modulation profile with uniform (space-time) slabs in the frequency domain\footnote{Note that, however, the uniform slab and frequency domain requirements of the transfer-matrix methods are very restrictive. The former requirement precludes the simulation of a 
space-time structures composed of slabs with \emph{finite temporal durations} (and hence invovling space-time corners), which are obviously a basic constraint in practical scenarios. The latter requirement essentially precludes the determination of the spatio-temporal features of the scattered waves because this would imply an extremely complex set of Fourier transformations. In contrast, the proposed generalized FDTD scheme can straightforwardly simulate
truncated slabs and exact space-time waveforms.}. This is even true for a smooth gradient modulation profile, which can be approximated as a succession of subwavelength interfaces with constant parameters. 

The transfer-matrix method~\cite{tsai1967wave,biancalana2007dynamics,deck2019uniform} models interfaces by the discontinuity matrix
\begin{equation}
    I_{mn}=\frac{1}{t_{nm}^-}\begin{bmatrix}
        t_{mn}t_{nm}^--\gamma_{mnm}\gamma_{nmn}^- & \gamma_{mnm}^-\\
        -\gamma_{nmn}&1
    \end{bmatrix},
\end{equation}
whose coefficients are given by 
\begin{equation}\label{eq:coefs_mn}
    t_{mn}=\frac{2\eta_m}{\eta_m+\eta_n}\frac{1-n_n v/c}{1-n_m v/c} \quad \text{and} \quad \gamma_{nmn}=\frac{\eta_m-\eta_n}{\eta_m+\eta_n}\frac{1-n_n v/c}{1-n_n v/c}
\end{equation}
and where the $t_{mn}^-$ and $\gamma_{nmn}^-$ coefficients are found by inverting the sign of $v$ in~\eqref{eq:coefs_mn} and the $t_{nm}$ and $\gamma_{nmn}$ coefficients are found by inverting $m$ and $n$ in~\eqref{eq:coefs_mn}. On the other hand, it models (space-time) slabs by the propagation matrix
\begin{equation}
    P_{m}=\begin{bmatrix}
        e^{-i\omega_m^+n_m\Delta z} & 0\\
        0&e^{i\omega_m^-n_m\Delta z}
    \end{bmatrix},
\end{equation}
where
\begin{equation}\label{eq:freq_grad}
    \omega_\tx{m}^+=\frac{1-n_1v/c}{1-n_\tx{m}v/c}\omega^+_1,
    \quad\tx{and}\quad
    \omega_\text{m}^-=\frac{1-n_1v/c}{1+n_m v/c}\omega_1^+
\end{equation}
are the frequencies of the forward- and backward-propagating waves within a given slab, and $\Delta z$ is the width of this slab (which is constant under the assumption of uniform discretization).

The transfer matrix of a structure composed of $N$ slabs is then found by taking the product of alternating discontinuity and propagation matrices, viz.,
\begin{equation}
    [M]_{N1}=[I_{N,N-1}][P_{N-1}][I_{N-1,N-2}][P_{N-2}]...[P_2][I_{2,1}].
\end{equation}

Finally, the scattering coefficients can be obtained by converting the transfer matrix into a scattering matrix~\cite{born1980principles}, which yields 
\begin{equation}\label{coef_matrix}
    T=\frac{A D-B C}{D}, \quad \Gamma=\frac{-A}{D},
\end{equation}
where $A$, $B$, $C$ and $D$ are the $(1,1)$, $(1,2)$, $(2,1)$ and $(2,2)$ components of the $[M]_{N1}$ matrix.

\bibliography{FDTD_modified_Yee.bib}
 
\end{document}